\documentclass[prd,twocolumn,amsmath,amssymb,floatfix,superscriptaddress,notitlepage,nofootinbib]{revtex4-1}

\usepackage{amsfonts,amssymb,amsmath,graphicx,color,bm}
\definecolor{ultramarine}{rgb}{0.07, 0.04, 0.56}
\definecolor{cadmiumgreen}{rgb}{0.0, 0.42, 0.24}
\definecolor{indigo(dye)}{rgb}{0.0, 0.25, 0.42}
\usepackage[linktocpage=true]{hyperref}
\hypersetup{
colorlinks=true,
citecolor=ultramarine,
linkcolor=cadmiumgreen,
urlcolor=indigo(dye),
}

\newcommand{\f}[2]{\frac{#1}{#2}}  
\newcommand{\mk}[1]{\left( #1 \right)}  
\newcommand{\kk}[1]{\left[ #1 \right]}  
  
\newcommand{\be}{\begin{equation}}  
\newcommand{\ee}{\end{equation}}
\newcommand{\Mpl}{M_{\rm Pl}}

\newcommand{\pa}{\partial}

\begin{document}

\title{  
$f(R)$ constant-roll inflation 
}

\author{Hayato Motohashi}
\affiliation{Instituto de F\'{i}sica Corpuscular (IFIC), Universidad de Valencia-CSIC, E-46980, Valencia, Spain}

\author{Alexei A.\ Starobinsky}
\affiliation{L. D. Landau Institute for Theoretical Physics RAS, Moscow, 119334 Russia}
\affiliation{National Research University Higher School of Economics, Moscow, 101000 Russia}

\begin{abstract}
The previously introduced class of two-parametric phenomenological inflationary models in General Relativity in which the slow-roll assumption is replaced by the more general, constant-roll condition is generalized to the case of $f(R)$ gravity. 
A simple constant-roll condition is defined in the original Jordan frame, and exact expressions for a scalaron potential in the Einstein frame,  for a function $f(R)$  (in the parametric form) and for inflationary dynamics are obtained.  The region of the model parameters permitted by the latest observational constraints on the scalar spectral index and the tensor-to-scalar ratio of primordial metric perturbations generated during inflation is determined.
\end{abstract}

\maketitle


\section{Introduction}
\label{sec:int}

The constant-roll inflation is a two-parametric class of phenomenological inflationary model which satisfies the assumption of constant rate of the inflaton~\cite{Martin:2012pe,Motohashi:2014ppa,Motohashi:2017aob}.  
The assumption is a generalization of the standard slow-roll inflation with an approximately flat inflaton potential, and so-called ultra-slow-roll inflation~\cite{Tsamis:2003px,Kinney:2005vj,Namjoo:2012aa}, in which the potential is constant for an extended period, and the curvature perturbation grows on superhorizon scales.  The attempt of such a generalization first proposed in \cite{Martin:2012pe}, and the inflaton potential was constructed so that it satisfies the constant-roll condition approximately.  Later, it was clarified in \cite{Motohashi:2014ppa} that there exists a potential that satisfies the constant-roll condition exactly.  In addition, the model possesses the exact solution that is an attractor for inflationary dynamics.  It is also elucidated that the curvature perturbation is conserved on superhorizon scales. Not only does the constant-roll inflation serve theoretically interesting framework, it is also viable with the most recent observational data. In \cite{Motohashi:2017aob}, we showed that the model can satisfy the latest observational constraint on the spectral index of the curvature power spectrum and the tensor-to-scalar ratio.

This constant-roll construction refers to inflationary models in General Relativity (GR) where gravity is not modified but a new scalar field has to be introduced. On the other hand, in the opposite limit one can construct inflationary models without new scalar fields, by changing the gravity sector only, as typified by the $R+R^2$ model \cite{Starobinsky:1980te} and its $f(R)$ gravity modifications~\cite{Barrow:1988xh,Appleby:2009uf,Motohashi:2012tt,Nishizawa:2014zra,Motohashi:2014tra}. This purely geometrical approach is equivalent to introducing a scalar degree of freedom (dubbed a scalaron in~\cite{Starobinsky:1980te}), which can be explicitly seen by performing a conformal transformation from the Jordan frame to the Einstein frame. Viable inflationary models in $f(R)$ gravity are slow-rolling, too. Since the present level of accuracy of astronomical observations make it interesting to go beyond the slow-roll approximation, in this paper we construct a new constant-roll inflationary model in the framework of $f(R)$ gravity. In contrast to the previous works~\cite{Martin:2012pe,Motohashi:2014ppa,Motohashi:2017aob} where the constant-roll condition was effectively imposed in the Einstein frame, since inflation in GR was considered, we impose a new constant-roll condition in the original Jordan frame where the form of equations is simpler in fact; see e.g.\ Eq.~(\ref{Jordan-eq}) below.

The rest of the paper is organized as follows.
In \S \ref{sec:frg}, we review $f(R)$ gravity focusing on its Jordan and Einstein frame description.
In \S \ref{sec:pot}, we introduce a novel constant-roll condition in the Jordan frame and derive exact solutions for the potential and the Hubble parameter in the Einstein frame.
In \S \ref{sec:con}, we derive a parametric expression of $f(R)$, and explore the inflationary dynamics in the Jordan frame. 
In \S \ref{sec:spe}, we consider the spectral parameters for the inflationary power spectra. 
We use them in \S \ref{sec:obs} to show the model possesses an available parameter region.
We conclude in \S \ref{sec:conc}.
In the Appendix, two alternative derivations of the parametric expression for the constant-roll $f(R)$ function are presented, with the latter of them using the Jordan frame only.

\section{$f(R)$ gravity}
\label{sec:frg}

Let us briefly review $f(R)$ gravity and the relation between the Einstein and Jordan frames (see e.g.\ \cite{DeFelice:2010aj} for a more extensive review and the
list of references).
We consider the action
\be S=\int d^4x \sqrt{-g_J} \f{f(R_J)}{2}, \ee
whose gravitational field equations 
for the Friedmann-Lema\^{i}tre-Robertson-Walker (FLRW) background with zero spatial curvature
\begin{align}
ds^2=-dt_J^2+a_J^2(t_J)(dx^2+dy^2+dz^2) 
\end{align}
and in the absence of other matter
are given by
\begin{align} \label{Jeqs}
3FH_J^2&=\f{1}{2}(R_JF-f)-3H_J\dot F , \notag\\
2F\dot H_J&=-\ddot F+H_J\dot F,
\end{align}
where $F\equiv df/dR$, the subscript $J$ denotes the Jordan frame, a dot denotes a derivative with respect to the Jordan frame time $t_J$, and we work in the unit where $\Mpl=(8\pi G)^{-1/2}=1$.  
By using the conformal transformation $g^E_{\mu\nu} = F g^J_{\mu\nu}$,
we can transform the gravitational kinetic term into the Einstein-Hilbert form.  
Further, we can normalize the scalar kinetic term as
\be S = \int d^4x \sqrt{-g_E} \kk{ R_E - \f{1}{2} (\pa_\mu \phi)^2 - V(\phi) }, \ee
where the subscript $E$ denotes the Einstein frame, and 
\begin{align} \label{FV}
F &= e^{\sqrt{\f{2}{3}} \phi}, \notag\\
V(\phi)&=\f{R_J F-f}{2F^2}.  
\end{align}
Once a functional form of $f(R)$ is specified in the Jordan frame, the scalaron $\phi$ and the potential $V(\phi)$ in the Einstein frame are given by the above definition.
Conversely, once the potential is specified in the Einstein frame, the Ricci scalar and the function $f(R)$ in the Jordan frame are given by
\begin{align}
R_J &= e^{\sqrt{\f{2}{3}} \phi} \mk{\sqrt{6} V_\phi + 4 V} , \notag\\
f(R_J) &= e^{2\sqrt{\f{2}{3}} \phi} \mk{\sqrt{6} V_\phi + 2 V} ,  \label{invtrans}
\end{align}
where $V_\phi\equiv \pa V/\pa\phi$.

The time coordinate and the scale factor in the Jordan and Einstein frame are related through
\begin{align}
dt_J &= e^{-\f{\phi}{\sqrt{6}}} dt_E, \notag\\
a_J &= e^{-\f{\phi}{\sqrt{6}}} a_E, 
\end{align}
from which we obtain the relation between Jordan frame quantities and the Einstein frame quantities:
\begin{align}
H_J &= e^{\f{\phi}{\sqrt{6}}} \mk{H_E - \f{1}{\sqrt{6}} \f{d\phi}{dt_E}} , \notag\\
\dot \phi &= e^{\f{\phi}{\sqrt{6}}} \f{d\phi}{dt_E}, \notag\\
\ddot \phi &= e^{\sqrt{\f{2}{3}} \phi } \kk{ \f{d^2\phi}{dt_E^2} + \f{1}{\sqrt{6}}\mk{\f{d\phi}{dt_E}}^2  } .
\end{align}
The Einstein equation and the Klein-Gordon equation in the Einstein frame 
for the FLRW background in the absence of the spatial curvature and other matter
take the standard form:
\begin{align}
&H_E^2 = \f{1}{3} \kk{ \f{1}{2}\mk{\f{d\phi}{dt_E}}^2 +V }, \notag\\
&\f{dH_E}{dt_E} = -\f{1}{2}\mk{\f{d\phi}{dt_E}}^2, \notag\\
&\f{d^2\phi}{dt_E^2} + 3H_E \f{d\phi}{dt_E} + \f{\partial V}{\partial \phi} =0 ,  \label{einE}
\end{align}
where $H_E \equiv \frac{1}{a_E} \frac{d a_E}{d t_E}$.

It is known that these equations can be reduced to one non-linear first-order differential equation for $H_E(\phi)$ of the Hamilton-Jacobi type~\cite{Muslimov:1990be,Salopek:1990jq}. However, for $f(R)$ gravity the master first-order equation for $H_J$ in the original Jordan frame considered as a function of the Ricci scalar $R_J$ has even a simpler form,
which can be obtained as follows.  
We represent $\dot F$ as
\be \dot F=\frac{dF(R_J)}{dR_J}\frac{dR_J}{dH_J}\dot H_J = \frac{dF(R_J)}{dR_J}\frac{dR_J}{dH_J}\frac{R_J-12H_J^2}{6} , \ee
and plug it to the last term of the first equation of \eqref{Jeqs} to obtain  
\be \label{Jordan-eq}
\frac{dH_J}{dR_J}=\frac{H_J(R_J-12H_J^2)}{(R_J-6H_J^2)F(R_J)-f(R_J)}\ \frac{dF(R_J)}{dR_J} . \ee
Note that the right-hand side can be explicitly written down once a functional form of $f(R_J)$ is specified, and hence \eqref{Jordan-eq} is the master first-order equation for $H_J$ as a function of $R_J$.

\section{$f(R)$ constant-roll potential}
\label{sec:pot}

In the previous works~\cite{Martin:2012pe,Motohashi:2014ppa,Motohashi:2017aob}, we considered the Einstein-Hilbert action with a 
canonical scalar field, and imposed the constant-roll condition $\ddot\phi = \beta H \dot\phi$.
Now we consider a natural generalization of the constant-roll condition in $f(R)$ gravity:
\be \ddot F = \beta H_J \dot F\, . \label{Jconroll} \ee 
As we shall confirm below, the slow-roll regime amounts to $\beta\to 0$, whereas a constant potential corresponds to $\beta\to -3$. Note that this condition is {\em not}
conformally dual to the former one used in GR. Of course, such generalization can be produced in many ways. We have chosen just the 
form \eqref{Jconroll} for the constant-roll condition in $f(R)$ gravity from reasons of simplicity and aesthetic elegance.~\footnote{When this paper
was prepared for submission, a paper on the same topic~\cite{Nojiri:2017qvx} has appeared in the archive. However, two different slow-roll conditions 
in $f(R)$ gravity proposed in that paper differ from the our one~\eqref{Jconroll} and lead to more complicated forms of $V(\phi)$ and $f(R)$.}  
In particular, in the case of the $R+R^2$ inflationary model, it reduces to
\be \ddot R_J = \beta H_J \dot R_J\, . \label{Rconroll} \ee
Note that, as we shall see below, while $R+R^2$ model does not have constant-roll solution, there exist constant-roll solutions for $R^p$ models.
Also, for a generic $f(R)$ function,
substituting the constant-roll condition \eqref{Jconroll} to \eqref{Jeqs} and integrating it, we obtain a very simple and elegant relation which has to be satisfied
for all models in this class at all times:
\be \label{FH} F(R_J)\propto H_J^{2/(1-\beta)}\, . \ee
After obtaining an analytic solution for Hubble parameter we can determine a proportionality constant.  We shall come back to this point soon.

Let us now find the corresponding effective potential for the dual representation of this model in the Einstein frame. In terms of the Einstein frame variables, the 
condition~\eqref{Jconroll} reads
\be \label{Econroll1} \f{d^2\phi}{dt_E^2} + \f{3+\beta}{\sqrt{6}} \mk{\f{d\phi}{dt_E} }^2  - \beta H_E \f{d\phi}{dt_E}  = 0 . \ee
Plugging this condition to the Klein-Gordon equation~\eqref{einE}, we obtain
\be (3+\beta) \kk{ H_E \f{d\phi}{dt_E} - \f{1}{\sqrt{6}} \mk{\f{d\phi}{dt_E} }^2 } + \f{\pa V}{\pa\phi} = 0 , \ee
where the quadratic velocity term shows up as we impose the constant-roll in the Jordan frame $\ddot F = \beta H_J \dot F$, rather than $d^2\phi/dt_E^2 = \beta H_E d\phi/dt_E$ in the Einstein frame.
Clearly, the limit $\beta\to -3$ amounts to the constant potential.
On the other hand, for the limit $\beta\to 0$, we have a slow-roll equation which is approximately equivalent to the standard form as the quadratic velocity term is negligible for slow roll.

Below we shall show that one can construct an inflationary model that satisfies the constant-roll condition~\eqref{Econroll1}, and has an exact solution for inflationary evolution.  Further, we shall clarify that the model has a parameter region that satisfies the latest observational constraint on spectral parameters of inflationary power spectra.

Following \cite{Motohashi:2014ppa}, we employ the Hamiltonian-Jacobi formalism and regard $H_E=H_E(\phi)$, assuming that $t_E = t_E(\phi)$ is a single-valued function, or $d\phi/dt_E\neq 0$.
When $d\phi/dt_E=0$, the Hamiltonian-Jacobi formalism breaks down, and the stochastic effect becomes dominant. 
It should be avoided that the inflaton passes such a point during inflation.
If the breakdown is located before inflation, there is no problem to rely on the Hamiltonian-Jacobi formalism.
We will check this point later on.

From the Einstein equation~\eqref{einE}, we obtain
\begin{align}
\f{d\phi}{dt_E} &= -2 \f{dH_E}{d\phi} ,  \notag\\
\f{d^2\phi}{dt_E^2} &= -2 \f{d^2H_E}{d\phi^2} \f{d\phi}{dt_E} ,  \label{dfdt}
\end{align}
with which the condition~\eqref{Econroll1} is rewritten as
\be \f{dH_E}{d\phi} \kk{ \f{d^2H_E}{d\phi^2} + \f{3+\beta}{\sqrt{6}}\f{dH_E}{d\phi} + \f{\beta}{2} H_E } = 0 . \ee

The equation allows two branches of solutions.
The first branch $dH_E/d\phi=0$ gives $H_E={\rm const}$.\ and $V={\rm const}$.\ in the Einstein frame, which corresponds to $f(R_J)=R_J-{\rm const}$.
In the second branch, the general solution is given by
\begin{align} \label{HEsol}
H_E(\phi) 
&= M ( \gamma e^{-\sqrt{\f{3}{2}} \phi } + e^{-\f{\beta\phi}{\sqrt{6}} } ) ,  
\end{align}
and the potential is given by
\begin{align}
V (\phi) &= 3H_E^2 - 2\mk{\f{dH_E}{d\phi}}^2 \notag\\
&= \f{3-\beta}{3} M^2 \kk{ 6 \gamma e^{-\f{(3+\beta)\phi}{\sqrt{6}} }  + (3+\beta) e^{-\sqrt{\f{2}{3}} \beta\phi  }  } \label{conpot}, 
\end{align}
where we introduced two integration constants $M$ (mass dimension 1) and $\gamma$ (dimensionless).
Using redefinition of $M$ and $\phi$, we can always normalize $\gamma$.  
Therefore, without loss of generality, we consider only $\gamma=0,\pm 1$ for the following.
On the other hand the amplitude of $M$ is determined by the CMB normalization.
Below we work in the unit where $M=1$.

\begin{figure}[t]
  \centering
  \includegraphics[width=\columnwidth]{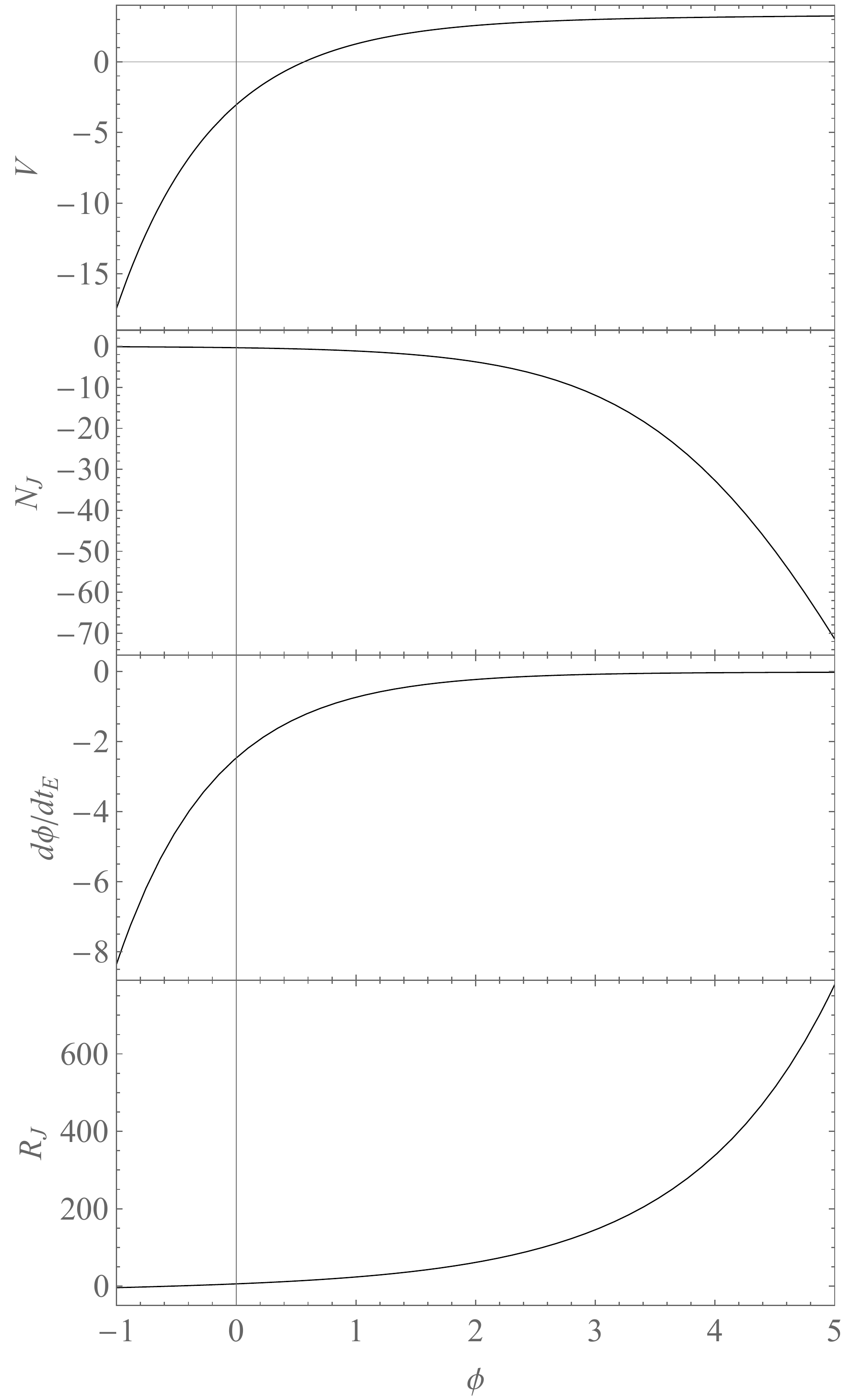}
  \caption{The potential~\eqref{conpot} in the Einstein frame, the phase diagram for the scalaron, the e-folds~\eqref{NJ} and the Ricci curvature~\eqref{RJfJ} in the Jordan frame for $\beta=-0.02, \gamma=-1$, where we set $\Mpl=M=1$.}
  \label{fig:pots}
\end{figure}

Below we shall clarify that viable parameter set is $\beta\lesssim 0$ and $\gamma = -1$.
Unlike the constant-roll potential found in \cite{Motohashi:2014ppa} using the condition $\ddot\phi=\beta H\dot\phi$, the potential \eqref{conpot} is not periodic function. 
Its form is depicted in Fig.~\ref{fig:pots} for a specific parameter set $\beta=-0.02$ with $\gamma=-1$.

Depending on values of the parameters $\beta$ and $\gamma$, the potential can pass $V=0$.
We can solve $V(\phi)=0$ and the solution is given by $\phi=\phi_c$, where we define the critical field value as 
\be \phi_c \equiv \f{\sqrt{6}}{\beta-3} \log \left| \f{3+\beta}{6\gamma} \right|  . \ee
For instance, $\phi_c\approx 0.57$ for $\beta=-0.02, \gamma=-1$.
A negative value of the potential is undesirable since it may lead to recollapse soon after the end of inflation. 
For this reason, we cut the potential at some point $\phi=\phi_0 > \phi_c$ to realize a graceful exit from inflation.

For $\gamma=0$, the potential is given by a single exponential function.
On the other hand, for $\beta\approx 0$ or $-3$ the potential is mainly described by a single exponential function with a constant, which is of our target.
However, we do not consider the case $\beta < -3$ and $\gamma = -1$ as the potential is always negative.

\section{$f(R)$ constant-roll dynamics}
\label{sec:con}

As mentioned above, from the form of the potential we focus on $\beta\approx 0$ or $-3$.
In this section we shall check inflationary dynamics. 
The evolution of the inflaton is governed by 
\be \f{d\phi}{dt_E} = -2 \f{dH_E}{d\phi} 
= \f{2}{\sqrt{6}} e^{-\f{\beta\phi}{\sqrt{6}} } ( 3 \gamma E + \beta  ) , \ee
where $E(\phi)\equiv e^{(\beta-3)\phi/\sqrt{6}}$.
By solving this equation, we obtain
\be t_E = \f{e^{\sqrt{\f{3}{2}}\phi} }{3\gamma} {}_2F_1 \mk{ 1,\f{3}{3-\beta},\f{6-\beta}{3-\beta}; -\f{\beta}{\gamma} e^{\f{(3-\beta)\phi}{\sqrt{6}}} } , \ee
where ${}_2F_1$ is the Gauss' hypergeometric function.
Thus, $\phi(t_E)$ is obtained in terms of the inverse function of the hypergeometric function.
However, without its specific form of the solution, we can draw interesting conclusion as follows.

If $\beta$ and $\gamma$ have the opposite sign, there exists a solution for $d\phi/dt_E \propto 3 \gamma E + \beta = 0$, which we denote $\phi_b$,
\be \phi_b \equiv \f{\sqrt{6}}{\beta-3} \log \left| \f{\beta}{3\gamma} \right| . \ee
We can show that if $\phi_b$ exists, the inflaton will always approach $\phi_b$ spending infinite Einstein-frame time as follows.
If $3 \gamma E + \beta > 0$, $\phi(t_E)$ is increasing and $3 \gamma E(t_E) + \beta$ is decreasing.
It continues decreasing so long as $3 \gamma E + \beta > 0$, and thus the inflaton approaches $\phi=\phi_b$.
Likewise, for the opposite case with $3 \gamma E + \beta < 0$, the inflaton also approaches $\phi=\phi_b$.
In both cases, the inflaton velocity $|d\phi/dt_E|$ is always decreasing, therefore it approaches $\phi=\phi_b$ spending infinite Einstein-frame time.

Actually, this process develops small-scale inhomogeneity of the Universe.
From the conformal invariance of the curvature perturbation, 
\be \zeta_J \sim \f{H_E}{d\phi/dt_E} \delta \phi \sim \f{H_E^2}{d\phi/dt_E}, \ee
where the right-hand side is evaluated at the horizon exit.
As we showed above, $|d\phi/dt_E|$ is always decreasing in the course of inflation.  
In such a case, $|\zeta_J|$ is amplified on small scales, and the Universe becomes inhomogeneous. 
This also means that the isotropic background solution involved is not an attractor. 
Therefore, we exclude the parameter set with $\beta \gamma < 0$ (see Table~\ref{tab:betagam}).

\begin{figure}[t]
  \centering
  \includegraphics[width=\columnwidth]{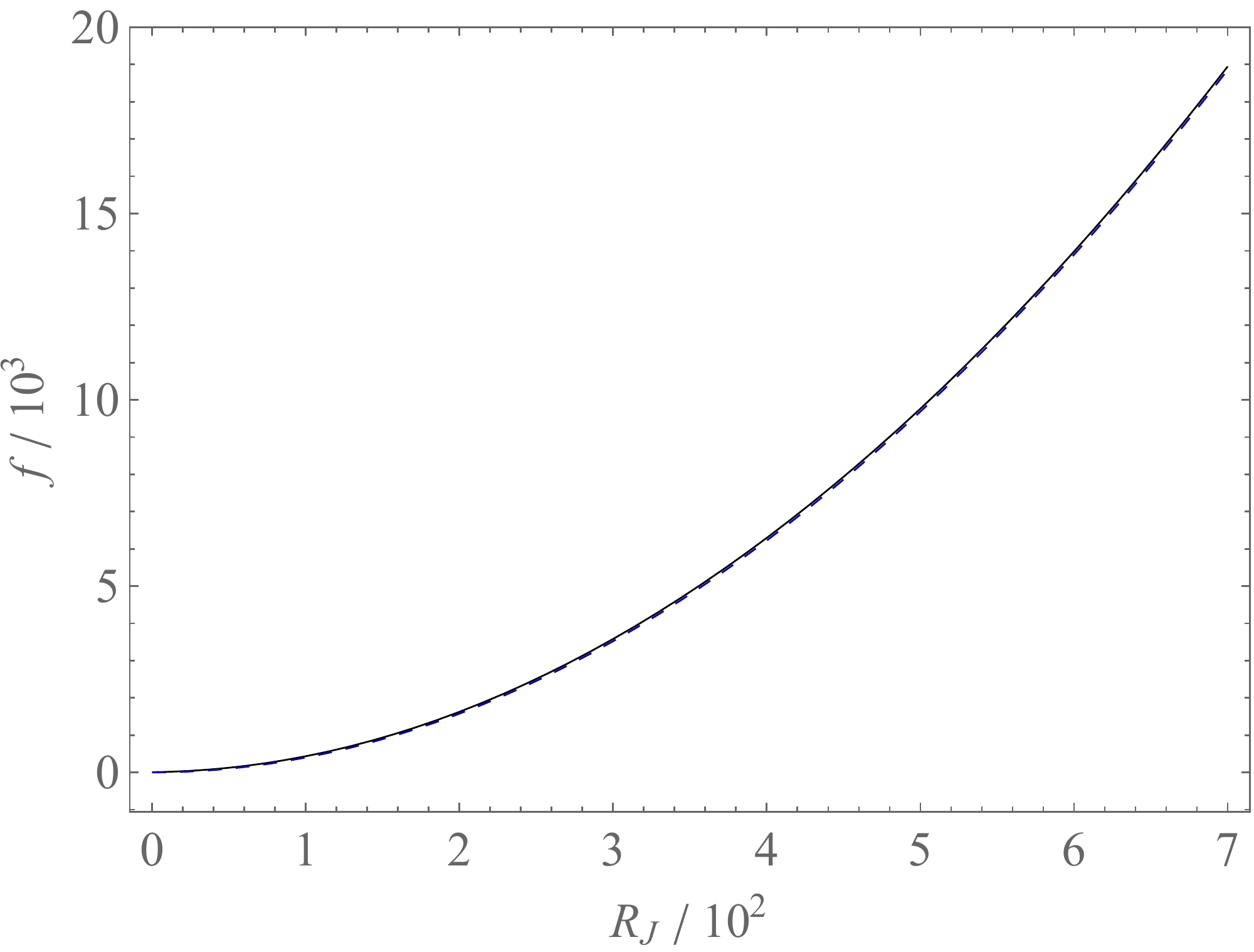}
  \caption{The function $f(R_J)$ for $\beta=-0.02, \gamma=-1$ in the exact parametric form~\eqref{RJfJ} (black, solid) and the approximated form~\eqref{fapp1} (blue, dashed). }
  \label{fig:fR}
\end{figure}

Plugging the potential~\eqref{conpot} to \eqref{invtrans}, we obtain a parametric expression of $f(R)$:
\begin{align}
&R_J = \f{2}{3}(\beta-3) e^{ 2(1-\beta)\phi/\sqrt{6} } \kk{ 3\gamma(\beta-1) E + (\beta-2)(\beta+3)  } \notag\\
&=(\beta-3)\kk{ 2\gamma(\beta-1)F^{-(1+\beta)/2}+ \frac{2}{3}(\beta-2)(\beta+3)F^{1-\beta}  }  ,   \notag\\
&f(R_J) = \f{2}{3}(\beta-3) e^{2(2-\beta)\phi/\sqrt{6} } \kk{ 3\gamma(\beta+1) E + (\beta-1)(\beta+3)  } \notag\\
&=(\beta-3) \kk{ 2\gamma(\beta+1)F^{(1-\beta)/2}+\frac{2}{3}(\beta-1)(\beta+3)F^{2-\beta} } .  
\label{RJfJ}
\end{align}
Here $F(\phi)=e^{\sqrt{\frac{2}{3}}\phi}$ serves as an auxiliary variable. However, it is easily seen that $df/dR=F$, as it should be.
For $\gamma=0$ or $\beta = -3$, we can write down $f(R_J)\propto R_J^p$, with $p=\f{2-\beta}{1-\beta}$ or $p=\f{\beta-1}{\beta+1}$, respectively.
For general case with $\beta \lesssim 0, \phi > 1$, the $R_J, f(R_J)$ in \eqref{RJfJ} are dominated by the second terms.  Neglecting the first terms, we obtain
\begin{align} \label{fapp1} 
f(R_J) &\approx \f{2}{3} (\beta - 3) (\beta + 3) (\beta - 1) \notag\\
&~~~~ \times \mk{\f{3R_J}{2 (\beta - 3) (\beta + 3) (\beta - 2)} }^{\f{2-\beta}{1-\beta}} . 
\end{align}
The high curvature behaviour is thus close to the $R+R^p$ model.  
Since it is shown in \cite{Motohashi:2014tra} that the $R+R^p$ model possesses a parameter region to satisfy the latest observational constraint, we expect that the present case would also be observationally viable.
Indeed, we shall see in \S\ref{sec:obs} that there exists a parameter region $\gamma=-1, - 0.1 \lesssim \beta \leq 0, 4 \leq \phi \leq 4.8$, which satisfies the latest observational constraint on inflationary power spectra. 
For this parameter region, we confirm that the relative error between the exact parametric form~\eqref{RJfJ} and the approximated form~\eqref{fapp1} remains less than $1.6\%$.
The exact and approximated forms of $f(R)$ are depicted in Fig.~\ref{fig:fR} for the case $\beta=-0.02, \gamma=-1$, for which $4 \leq \phi \leq 4.8$ amounts to $3.4 \leq R_J/10^2 \leq 6.6$. 
The relative error increases as $\phi$ or $R_J$ decreases, and reaches $5,10\%$ at $R_J/10^{-2}=1.4$, $0.88$, respectively.

Since in the inflationary regime Ricci curvature in \eqref{RJfJ} should be positive,
we are interested in the field region that satisfies 
\be (\beta-3)[3\gamma(\beta-1) E + (\beta+3)(\beta-2)] > 0. \ee 
However, for the case $\beta < -3$ and $\gamma = -1$, Ricci curvature is always negative, which is another reason why we do not consider this parameter set, in addition to the negative potential mentioned above.
For other parameter sets, the Ricci curvature can pass $R_J=0$ and change the sign at $\phi=\phi_r$, where
\be \label{phir} \phi_r \equiv \f{\sqrt{6}}{\beta-3} \log \left| \f{(\beta+3)(\beta-2)}{3\gamma(\beta-1)} \right| . \ee
For instance, $\phi_r\approx -0.55$ for $\beta=-0.02, \gamma=-1$.  While Ricci curvature is negative for $\phi<\phi_r$, in this case it does not occur during inflation as we cut the potential at some point $\phi=\phi_0 > \phi_c \approx 0.57$ to realize a graceful exit from inflation.
For later convenience, it is worthwhile to note that the tensor-to-scalar ratio $r(\phi)$ at $\phi=\phi_r$ does not depend on $\beta$ nor $\gamma$ and takes very large value which is unacceptable from observational point of view:
\be \label{rR0} r|_{R_J=0} = \f{64}{3} \approx 21.3\, . \ee 
In addition, we require that the Ricci curvature is decreasing during inflation, namely,
\begin{align} \label{dRdt}
\f{dR_J}{dt_E} &= - 2 \f{dR_J}{d\phi} \f{dH_E}{d\phi} \notag\\
&= - \f{2}{9} e^{ \f{(2-3\beta)\phi}{\sqrt{6}} } (3 - \beta) (1 - \beta) (\beta + 3 \gamma E ) \notag\\
&~~~~ \times [ 2  (-2 + \beta) (3 + \beta) + 3 (1 + \beta) \gamma E ] , 
\end{align}
should be negative.
We use these expressions in \S\ref{sec:spe} to constrain the parameter space.

On the other hand, Hubble parameter in the Jordan frame is given by 
\be H_J = e^{\f{\phi}{\sqrt{6}}} \mk{ H_E + \f{2}{\sqrt{6}} \f{dH_E}{d\phi} } = \f{3-\beta}{3} e^{\f{(1-\beta)\phi}{\sqrt{6}}} . \ee
Using the definition $F = e^{ \sqrt{\f{2}{3}} \phi}$ we obtain
\be \label{FH2} F = \mk{\f{3 H_J}{3-\beta} }^{2 / (1-\beta)} ,\ee
which precisely reproduces \eqref{FH}.  Also, using the relation
\be H_J = \f{d N_J}{dt_J} = \f{dt_E}{dt_J} \f{d\phi}{dt_E} \f{dN_J}{d\phi},  \ee
we obtain e-folds in the Jordan frame as
\be \label{NJ} N_J = \f{1}{\beta} \log \left| 1 + \f{\beta}{3 \gamma} e^{\f{(3-\beta)\phi}{\sqrt{6}}} \right| . \ee

\section{Inflationary power spectra}
\label{sec:spe}

\begin{table*}[t] 
\centering
\caption{Reasons why parameter regions except $\beta\lesssim 0, \gamma=-1$ are excluded. }
\label{tab:betagam}
\begin{tabular}{cccc}\hline
 & $\gamma=+1$ & $\gamma=-1$ & $\gamma=0$ \\ \hline 
$\beta \lesssim -3$ & Inhomogeneity & Always $R,V<0$ & $r = 8 (1-n_s)$ \\ 
$\beta \gtrsim -3$ & Inhomogeneity & $r\geq 21.3$ for $R_J\geq 0$ & $r = 8 (1-n_s)$ \\ 
$\beta \lesssim 0$ & Inhomogeneity & Viable, Fig.~\ref{fig:conparams} & $r = 8 (1-n_s)$  \\ 
$\beta \gtrsim 0$  & $r\geq 9.48$ for $\f{dR_J}{dt_E}\leq 0$ & Inhomogeneity & $r = 8 (1-n_s)$ \\ \hline
\end{tabular}
\end{table*} 

Now we check the spectral parameters of inflationary power spectra and compare them with observational constraint to find viable parameter set $(\beta,\gamma)$. 
First, the power spectrum of scalar (curvature) and tensor perturbations can be calculated in the Jordan frame directly, e.g.\ as was quantitatively correctly done in \cite{Starobinsky:1983zz} for the model \cite{Starobinsky:1980te}
using the $\delta N$ formalism. 
Second, the calculation in the Einstein frame leads to the same result since the constant modes of scalar (curvature) and tensor perturbations are not affected by a generic (inhomogeneous) conformal transformation after the
end of inflation; see e.g.\ \cite{Chiba:2008ia,Gong:2011qe} for more details, and \cite{Motohashi:2015pra} for more general invariance under disformal transformation. The subtle point is
that though the value of the power spectrum is the same in both frames,
it refers to slightly different inverse scales $k_E$ and $k_J$. 
However, corrections to the power spectra of scalar and tensor
perturbations following from this difference are proportional to $|n_s-1|$ and $|n_t|$ correspondingly. In particular, they would be absent for the
exactly scale-invariant spectra. Thus, they can be neglected in the leading order of the slow-roll approximation.

We evaluate the spectral parameters by exploiting the slow-roll parameters for the inflaton potential in the Einstein frame, which are given by
\begin{align}
\epsilon &\equiv \f{1}{2} \mk{\f{V'}{V}}^2 
= \f{(3 + \beta)^2 (\beta + 3 \gamma E)^2}{3 (3 + \beta + 6 \gamma E)^2}, \notag\\ 
\eta&\equiv \f{V''}{V} = \f{(3 + \beta) [2 \beta^2 + 3 (3 + \beta) \gamma E]}{3 (3 + \beta + 6 \gamma E)} , \\
\xi&\equiv \f{V'V'''}{V^2} = \f{ (3 + \beta)^2 ( \beta + 3 \gamma E) [4 \beta^3 + 3 (3 + \beta)^2 \gamma E]}{ 9 (3 + \beta + 6 \gamma E)^2} , \notag
\end{align}
where a prime denotes derivative with respect to $\phi$.
We are interested in regime where these slow-roll parameters are sufficiently small to obtain a nearly scale-invariant spectrum.
By virtue of the conformal invariance of the curvature and tensor perturbations, we can use the standard slow-roll expansion of the spectral parameters 
\begin{align}
&n_s-1 = -6\epsilon+2\eta \notag\\
&~= -\f{ 2 (3 + \beta) }{ 3 (3 + \beta + 6 \gamma E)^2 } [\beta^2 (3 + \beta) \notag\\
&~\quad  + 3 (-9 + 12 \beta + \beta^2) \gamma E + 9 (3 + \beta) \gamma^2 E^2],\\
&r= 16\epsilon \notag\\
&~ = \f{16(3 + \beta)^2 (\beta + 3 \gamma E)^2}{3 (3 + \beta + 6 \gamma E)^2},\\
&\f{dn_s}{d\ln k}= 16\epsilon \eta-24\epsilon^2-2\xi \notag\\
&= \f{2 (9 - \beta^2)^2 \gamma E (\beta + 
   3 \gamma E) [(-1 + \beta) (3 + \beta) + 
   12 \gamma E]}{(3 + \beta + 6 \gamma E)^4} .
\end{align} 
For $\gamma=0$, they read 
\be n_s-1 \to -\f{2}{3}\beta^2 , \quad 
r\to \f{16}{3}\beta^2, \quad 
\f{dn_s}{d\ln k} \to 0. \ee
We thus obtain a consistency relation $r = 8 (1-n_s)$, for which it is impossible to satisfy the observational constraint.  
For instance, $r=0.32,0.24$ for $n_s=0.96,0.97$, respectively.
Therefore, below we focus on $\gamma=\pm 1$.

Finally, let us focus on the tensor-to-scalar ratio $r$ along with the evolution of $R_J$.  
As we mentioned in \S\ref{sec:con}, we require $R_J>0$ and $dR_J/dt_E<0$ during inflation.
For the parameter set $\beta\gtrsim -3$ with $\gamma=-1$, $R_J \geq 0$ for $\phi \geq \phi_r$ where $\phi_r$ is defined by \eqref{phir}. 
Then it can be shown by checking $dr/d\phi$ and $d^2r/d\phi^2$ that for the region $\phi \geq \phi_r$ the minimum value of $r$ is given at $\phi=\phi_r$ which is given by \eqref{rR0}.
Therefore, so long as we consider the region where $R_J \geq 0$, we have $r\geq 21.3$, which is much larger than the observationally allowed value.
Likewise, for the parameter set $\beta\gtrsim 0$ with $\gamma>0$, by using \eqref{dRdt} we can also show that $dR_J/dt_E\leq 0$ holds for $\phi \leq \f{\sqrt{6}}{\beta-3} \log \kk{ \f{2(2-\beta)(3+\beta)}{3\gamma(1+\beta)} }$, and for this field position, $r\geq \f{16}{27} (4+\beta)^2 \geq 9.48$, which is also not acceptable.

We thus find that only allowed possibility is $\beta \lesssim 0$ with $\gamma=-1$.
Indeed, parameter set $-0.02 \lesssim \beta < 0 ,\gamma=-1$ satisfies the observational constraint on $(n_s,r)$.
Other parameter regions are not feasible for various reasons, which are summarized in Table~\ref{tab:betagam}.

\begin{figure}[h]
  \centering
  \includegraphics[width=\columnwidth]{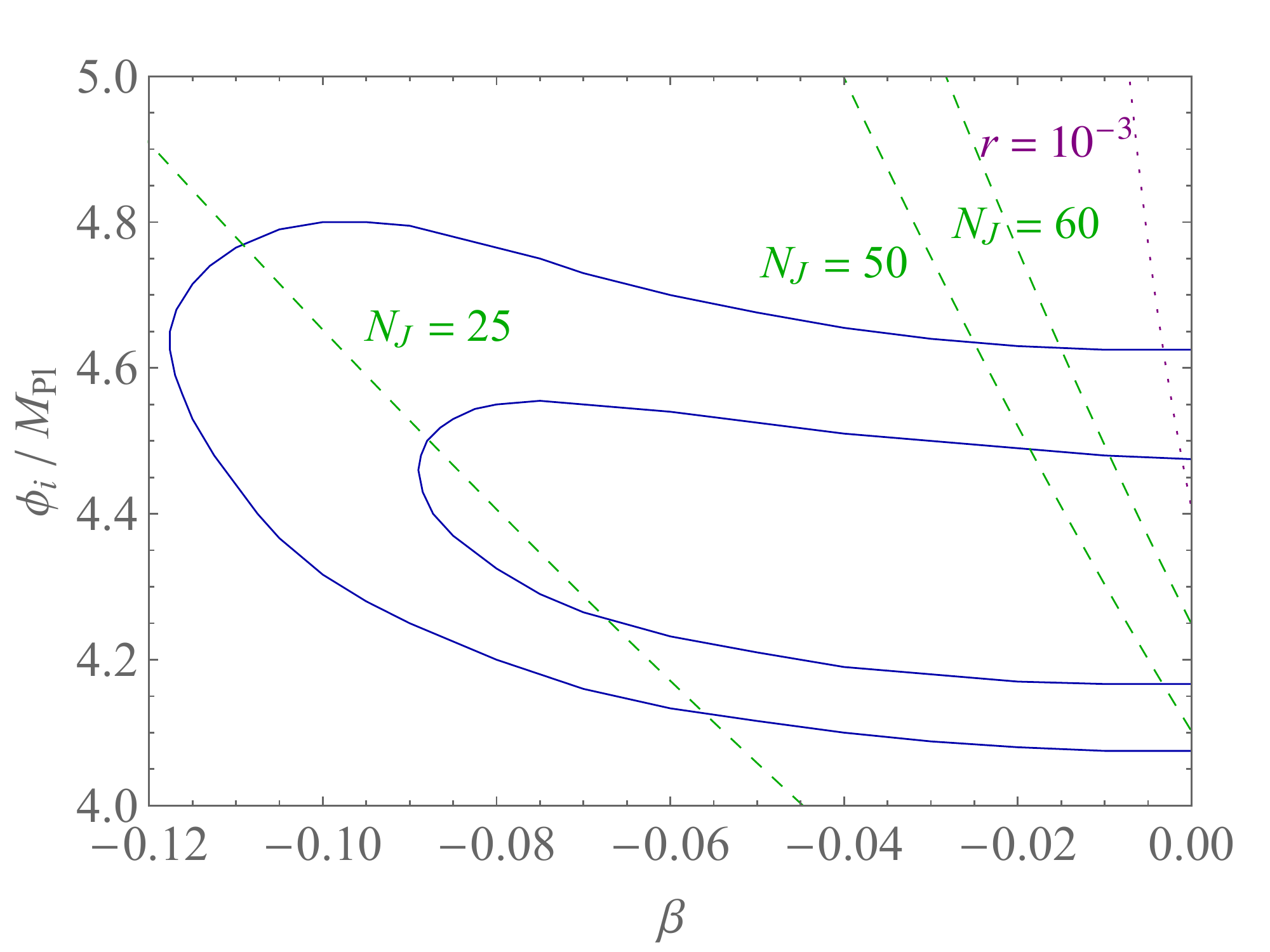}
  \caption{Observational constraint from CMB by Planck and BICEP2/Keck Array \cite{Array:2015xqh} on parameter space $(\beta, \phi_i/\Mpl)$ for the case $\beta\lesssim 0, \gamma=-1$.  The 68\% and 95\% confidence regions (blue), the Jordan frame e-folds $N_J$ counted back from $\phi_c$ (green, dashed), and $r=10^{-3}$ (purple, dotted). }
  \label{fig:conparams}
\end{figure}

\section{Observational constraints}
\label{sec:obs}

As shown in Table~\ref{tab:betagam}, in the previous sections we checked that the parameter regions other than $\beta \lesssim 0, \gamma=-1$ are excluded by various reasons. 
Now we show that the case $\beta \lesssim 0, \gamma=-1$ can indeed satisfy the latest observational constraint.

We already see the typical behavior for this parameter set in Fig.~\ref{fig:pots}.
The potential is approximated by $V \sim - e^{-\phi} + {\rm const}$ so long as $|\beta\phi|\lesssim 1$. 
The inflaton rolls on the plateau of the potential at positive $\phi$ region towards negative direction with $d\phi/dt_E < 0$.
Before the inflaton reaches to $\phi = \phi_c$ where $V=0$, we need to cut the potential at some point $\phi=\phi_0 > \phi_c$ to realize a graceful exit from inflation.

Indeed, this form of the potential with a long plateau is favored by the observational data.
During the inflation on the plateau, $V$ and $R_J$ remain positive, and the plateau is sufficiently long to produce a large number of e-folds $N_J\sim 50$ (see Fig.~\ref{fig:pots}).

Furthermore, we compare our model with the latest observational constraint on $(n_s,r)$ by Planck and BICEP2/Keck Array (see Fig.~7 in \cite{Array:2015xqh}).
Figure~\ref{fig:conparams} depicts the allowed parameter region.
We find that the parameter set $-0.02 \lesssim \beta < 0 ,\gamma=-1$ satisfies the observational constraint and provides a sufficient number of e-folds in the Jordan frame.
On the other hand, if the tensor-to-scalar ratio is close to the upper bound $r\sim 0.1$, this model has small number of e-folds $N_J\sim 25$.

\section{Conclusion}
\label{sec:conc}

We have constructed a simple and natural generalization of the class of constant-roll inflationary models in GR to the case of $f(R)$ gravity.  
The constant-roll condition ({\ref{Jconroll}) is introduced in the original Jordan frame. Using it, we derived the exact solutions for the Einstein-frame potential in \eqref{conpot}, the parametric expression of $f(R_J)$ in \eqref{RJfJ}, as well as the inflationary evolution in the Einstein and Jordan frames.
The functional form of $f(R_J)$ is expressed parametrically, while for some special parameter values it is possible to write down $f(R_J)$ explicitly as a function of $R_J$.
We showed that the model has an interesting parameter region $-0.1 \lesssim \beta < 0, \gamma=-1$ which satisfies the latest observational constraint on $(n_s,r)$ obtained by the Planck and BICEP2/Keck Array Collaborations.

\begin{acknowledgements}
H.M.\ thanks the Research Center for the Early Universe (RESCEU), where part of this work was completed.  He was also supported in part by 
MINECO Grant SEV-2014-0398,
PROMETEO II/2014/050,
Spanish Grant FPA2014-57816-P of the MINECO, and
European Union’s Horizon 2020 research and innovation programme under the Marie Sk\l{}odowska-Curie grant agreements No.~690575 and 674896.
A.S.\ acknowledges RESCEU for hospitality as a visiting professor. He was also partially supported by the grant RFBR 17-02-01008 and by the 
Scientific Programme P-7 (sub-programme 7B) of the Presidium of the Russian Academy of Sciences. 
\end{acknowledgements}

\appendix

\section{Alternative derivations of $f(R)$}
\label{sec:app}

In this appendix we present two alternative derivations for the parametric form of $f(R)$ in \eqref{RJfJ}, with the latter of them using the Jordan frame only.
First, we show an alternative derivation based on the results obtained in the Einstein frame.
By plugging \eqref{conpot} to the definition \eqref{FV}, 
we obtain a differential equation for $f(R)$ as 
\be \label{fRdiff}  f =  R f' + \f{2}{3}(\beta-3) f'^2 \kk{ 6\gamma f'^{-\f{3+\beta}{2}}  + (3+\beta) f'^{-\beta} } ,  \ee
where a prime denotes derivative with respect to $R$.
This equation is known as Clairaut's equation.

In general, Clairaut's equation is defined as
\be y(x) = x y' + g (y') , \ee
where a prime denotes derivative with respect to $x$.
By taking a derivative of the equation, we obtain
\be ( x + g' ) y'' = 0 , \ee
which clearly has two branches of solutions.
The first branch $y''=0$ yields general solution 
\be y = c x + g(c) , \ee
where $c$ is integration constant.  The second branch $x + g' = 0$ yields singular solution
\be x = - g'(c), \quad y = - c g'(c) + g(c) . \ee
This solution defines an envelope of the general solutions and thus does not involve any integration constant.  Thus $c$ plays a role of parameter and we have
\be \label{dydxc} \f{dy}{dx} = c . \ee
For some special case, $y$ can be written down explicitly as a function of $x$. 
For instance, if $g(c)=c^2$, the singular solution yields $x=-2c$ and $y = - c^2 = - x^2/4$.

In the present case, we have
\begin{align} 
g(c) &= \f{2}{3}(\beta-3) \kk{ 6\gamma c^{\f{1-\beta}{2}}  + (3+\beta) c^{2-\beta} } , \\
g'(c) &= \f{2}{3}(\beta-3) \kk{ 3(1-\beta) \gamma c^{-\f{1+\beta}{2}}  + (3+\beta)(2-\beta) c^{1-\beta} } , \notag
\end{align}
and the singular solution 
\be R = - g'(c), \quad f = - c g'(c) + g(c) , \ee
precisely reproduces \eqref{RJfJ} with $c=e^{\sqrt{\f{2}{3}}\phi} = df/dR$, as expected from \eqref{dydxc}.

Now let us show how to derive Eq.~\eqref{RJfJ} for the constant-roll $f(R)$ function working in the Jordan frame only.
Let us first denote the arbitrary constant appearing in \eqref{FH} as $A$:
\be F(R)=AH^{2/(1-\beta)}\, . \label{FHA} \ee
Introducing \eqref{FHA} into \eqref{Jordan-eq}, we get the following equation
\be f=\frac{\beta+1}{\beta-1}RF-\frac{6(\beta+3)}{\beta-1}\frac{F^{2-\beta}}{A^{1-\beta}} \, , \label{Jordan-eq1} \ee
which is a particular case of d'Alembert's differential equation (also known as Lagrange-d'Alembert's equation)
\be y(x)=xh(y')+g(y') \, , \label{Alembert} \ee
where the prime means derivative with respect to $x$.

The d'Alembert's equation is more general than  the Clairaut's equation considered above, and it can be solved using the same trick: considering $y'\equiv z$ as an independent variable and differentiating \eqref{Alembert} with respect to $z$ leads to the linear differential equation for $x=x(z)$:
\be \frac{dx}{dz}-\frac{xh'(z)}{z-h(z)}=\frac{g'(z)}{z-h(z)} \, . \ee
Now the prime means derivative of the functions $f$ and $g$ with respect to their argument $z$. Its general solution is
\begin{align} 
&x(z)=\exp\left(\int^z\frac{h'(u)}{u-h(u)}du\right)  \\
&\times\left[B+\int^z\frac{g'(u)}{u-h(u)} 
\exp \left(-\int^{u}\frac{h'(w)}{w-h(w)}dw\right) du\right]\, ,  \notag \label{solution} 
\end{align}
where $B$ is an integration constant.
In our case of \eqref{Jordan-eq1}, \eqref{solution} and \eqref{Jordan-eq1} take the form:
\begin{align} 
R&=\frac{6(\beta-2)(\beta+3)}{(\beta-3)A^{1-\beta}}F^{1-\beta}+BF^{-(1+\beta)/2} \, ,\\
f(R)&=\frac{6(\beta-1)(\beta+3)}{(\beta-3)A^{1-\beta}}F^{2-\beta}+B\frac{\beta+1}{\beta-1}F^{(1-\beta)/2} \, . \notag
\end{align}
This just coincides with \eqref{RJfJ} with $A^{1-\beta}=9/(\beta-3)^2$ from \eqref{FH2} (for $M=1$) and $B=2\gamma(\beta-3)(\beta-1)$.

\bibliography{ref-fRconroll}

\end{document}